\documentclass[11pt]{article}
\usepackage{graphicx}
\usepackage{amsmath}
\usepackage{amssymb}
\usepackage{amsfonts}
\usepackage{bbold}
\usepackage{geometry}

\newcommand\pubnumber{CP3-Origins-2017-017 DNRF90}
\newcommand\pubdate{\today}

\def\CP3{CP$^3$-Origins, University of Southern Denmark,\\
Campusvej 55, DK-5230 Odense M, Denmark}
\def\support{\footnote{The CP$^3$-Origins center is partially funded by the Danish National Research Foundation, grant number DNRF90.}}

\def\Title#1{\begin{center} {\Large #1 } \end{center}}
\def\Author#1{\begin{center}{ \sc #1} \end{center}}
\def\Address#1{\begin{center}{ \it #1} \end{center}}

\newcommand\pubblock{\rightline{\begin{tabular}{l} \pubnumber\\
         \pubdate  \end{tabular}}}
\newenvironment{Abstract}{\begin{quotation}  }{\end{quotation}}
\newenvironment{Presented}{\begin{quotation} \begin{center} 
             PRESENTED AT\end{center}\bigskip 
      \begin{center}\begin{large}}{\end{large}\end{center} \end{quotation}}
\def\Acknowledgements{\bigskip  \bigskip \begin{center} \begin{large}
             \bf ACKNOWLEDGEMENTS \end{large}\end{center}}

\def\beq{\begin{equation}}
\def\eeq#1{\label{#1}\end{equation}}
\def\eeqn{\end{equation}}

\def\beqa{\begin{eqnarray}}
\def\eeqa#1{\label{#1}\end{eqnarray}}
\def\eeqan{\end{eqnarray}}

\let\bar=\overbar

\def\Dslash{\not{\hbox{\kern-4pt $D$}}}
\def\dslash{\not{\hbox{\kern-2pt $\del$}}}

\def\msb{{\bar{\ssstyle M \kern -1pt S}}}

\begin{document}
\begin{titlepage}
\pubblock

\vfill
\Title{Theories of Leptonic Flavor}
\vfill
\Author{Claudia Hagedorn\support}
\Address{\CP3}
\vfill
\begin{Abstract}
I discuss different theories of leptonic flavor and their capability of describing the features of 
the lepton sector, namely charged lepton masses, neutrino masses, lepton mixing angles
and leptonic (low and high energy) CP phases. In particular, I show examples of theories with an abelian flavor
symmetry $G_f$, with a non-abelian $G_f$ as well as theories with non-abelian $G_f$ and CP.
\end{Abstract}
\vfill
\begin{Presented}
NuPhys2016: Prospects in Neutrino Physics,\\
Barbican Centre, London, UK,  December 12--14, 2016
\end{Presented}
\vfill
\end{titlepage}
\def\thefootnote{\fnsymbol{footnote}}
\setcounter{footnote}{0}
%

\section{Introduction}

Charged lepton masses have been determined with high precision~\cite{pdg16}
\begin{eqnarray}\nonumber
m_e&=& \left( 0.5109989461 \pm 0.0000000031 \right) \, \mbox{MeV} \, ,\\ \nonumber
m_\mu &=& \left( 105.6583745 \pm 0.0000024 \right) \, \mbox{MeV} \, ,\\ 
m_\tau &=& \left( 1776.86 \pm 0.12 \right) \, \mbox{MeV}
\end{eqnarray}
and turned out to be strongly hierarchical. In contrast, the neutrino mass spectrum is not yet completely known, but
the solar and the atmospheric mass squared differences $\Delta m_{\rm sol}^2$ and  $|\Delta m_{\rm atm}^2|$ have been measured in neutrino oscillation
experiments~\cite{nufit}
\begin{equation}
7.03\, \times 10^{-5}\, \mathrm{eV}^2 \lesssim \Delta m_{\rm sol}^2 \lesssim 8.09 \, \times 10^{-5}\, \mathrm{eV}^2
\;, \;\;
2.41 \, \times \, 10^{-3} \, \mathrm{eV}^2 \lesssim |\Delta m_{\rm atm}^2| \lesssim 2.64 \, \times \, 10^{-3} \, \mathrm{eV}^2
\end{equation}
and an upper bound on the sum of the neutrino masses of less than $1 \, \mathrm{eV}$ is derived from cosmology and beta decay experiments. 
The $3 \, \sigma$ ranges of the three lepton mixing angles, obtained in global fits of the data from neutrino experiments,
are~\cite{nufit}
\begin{equation}
0.01934 \lesssim \sin^2 \theta_{13} \lesssim 0.02397 \; , \;\;
0.271 \lesssim \sin^2 \theta_{12} \lesssim 0.345 \; , \;\;
0.385 \lesssim \sin^2 \theta_{23} \lesssim 0.638 \; .
\end{equation}
The nature of neutrinos, i.e. whether they are Dirac or Majorana particles, is still unknown. The following discussion assumes them to be Majorana particles. However,
most of the results also hold for neutrinos being Dirac particles. For now only hints exist for CP violation in the lepton sector~\cite{nufit}.

The strong hierarchy among the charged lepton masses, that is observed among the masses of the up and down type quarks as well, has led
to the consideration of abelian flavor symmetries $G_f$, usually called Froggatt-Nielsen (FN) symmetry $U(1)_{\mathrm{FN}}$~\cite{FN}.
The different generations of charged
leptons carry different charges under $U(1)_{\mathrm{FN}}$. As $\sqrt{\Delta m_{\mathrm{sol}}^2/|\Delta m_{\mathrm{atm}}^2|}\sim 1/6$, 
neutrino masses exhibit no strong hierarchy and it is thus expected that the neutrino sector is (partly) uncharged under $U(1)_{\mathrm{FN}}$ (see e.g.~below the assignment called leptonic anarchy). 
A disadvantage is that an FN symmetry is only capable to explain the order of magnitude of observables in terms of 
the (small) symmetry breaking parameter $\lambda$. 

For a non-abelian $G_f$ many choices of symmetries are available: if $G_f$ should be continuous, potentially suitable choices are $SO(3)$, $SU(2)$ and $SU(3)$,\footnote{An example of a model with a continuous non-abelian $G_f$ is found in~\cite{KingRoss}.} while for $G_f$ being discrete
indeed an infinite number of potentially suitable choices is known, like the series of dihedral groups $D_n$ ($n>2$), alternating groups $A_n$ ($n=4,5$), symmetric groups $S_n$ ($n=3,4$), the series $\Delta (3 \, n^2)$ and $\Delta (6 \, n^2)$ for $n>1$. An advantage of such non-abelian $G_f$ is the possibility to unify the three lepton generations partially, $L_\alpha \sim {\bf 2} +{\bf 1}$, or fully, $L_\alpha \sim {\bf 3}$. Furthermore, if broken to non-trivial residual symmetries~\cite{Gfresidual}, like $G_f=S_4$ that is broken to $Z_3$ in the charged lepton and to $Z_2 \times Z_2$ in the neutrino sector,  certain values of the lepton mixing parameters can be predicted, e.g. tri-bi-maximal mixing. 
Compared to an abelian $G_f$, model building with a non-abelian $G_f$ is, however, more challenging, e.g. more fields are needed, the construction of the potential in order to achieve the correct symmetry breaking pattern is non-trivial. 

Recently, theories with a discrete non-abelian $G_f$ have been extended with a CP symmetry~\cite{GfCPidea}, in particular, in order to also predict the Majorana phases $\alpha$ and $\beta$ and in order to obtain non-trivial values for the Dirac phase $\delta$. If such a theory comprises right-handed (RH) neutrinos, it is possible to generate the baryon asymmetry $Y_B$ of the Universe via (unflavored) leptogenesis. In this case the sign of $Y_B$ can be directly correlated with the results for the low energy CP phases $\alpha$, $\beta$ and $\delta$ ~\cite{GfCPlepto}.

\mathversion{bold}
\section{Theories with abelian $G_f$}
\mathversion{normal}

\begin{figure}
\parbox{6in}{\includegraphics{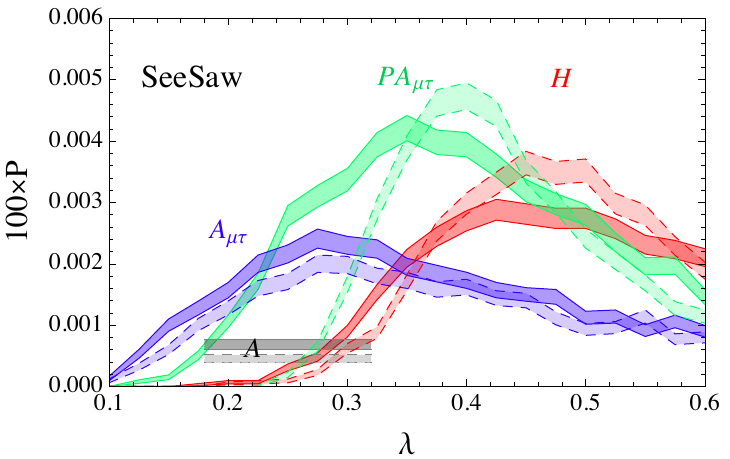}\hspace{0.5in}
\includegraphics{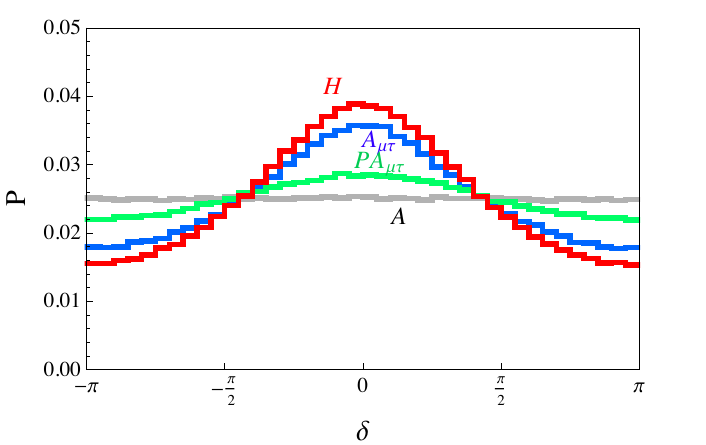}}
\caption{The left plot shows the success of the different charge assignments in describing correctly the lepton masses and mixing angles with respect to $\lambda$, assuming the seesaw mechanism responsible for neutrino masses. The right plot shows the probability distribution of the values of $\delta$ for a fixed value of $\lambda$, $\lambda=0.2$ for $A$, $A_{\mu\tau}$
and $\lambda=0.4$ for $H$ and $PA_{\mu\tau}$. For further details see~\cite{Altarelli12}. \label{fig1}}
\end{figure}

In~\cite{Altarelli12} the following charge assignments of the three generations of the left-handed (LH) lepton doublets $L_\alpha$, RH charged leptons $\alpha^c_L$
and RH neutrinos $\nu^c_i$ have been analyzed
\begin{eqnarray}\nonumber
\mathrm{leptonic} \, \mathrm{anarchy}\; (A)&:&\;\; L_\alpha \sim (0,0,0) \; , \;\; \alpha^c_L \sim (3,2,0) \; , \;\; \nu^c_i \sim (0,0,0) \; ,
\\ \nonumber
\mu\tau \mathrm{-anarchy}\; (A_{\mu\tau})&:&\;\; L_\alpha \sim (1,0,0) \; , \;\; \alpha^c_L \sim (3,2,0) \; , \;\; \nu^c_i \sim (2,1,0) \; ,
\\ \nonumber
\mathrm{pseudo} \, \mu\tau \mathrm{-anarchy}\; (PA_{\mu\tau})&:&\;\;  L_\alpha \sim (2,0,0) \; , \;\; \alpha^c_L \sim (5,3,0) \; , \;\; \nu^c_i \sim (1,-1,0) \; ,
\\ \nonumber
\mathrm{hierarchy}\; (H)&:&\;\; L_\alpha \sim (2,1,0) \; , \;\; \alpha^c_L \sim (5,3,0) \; , \;\; \nu^c_i \sim (2,1,0) \; .
\end{eqnarray}
The structure of the charged lepton mass matrix $m_l$ and the light neutrino mass matrix $m_\nu$, arising from leptonic anarchy, is 
\begin{equation}
m_l \sim \left( \begin{array}{ccc}
 \lambda^3 & \lambda^2 & 1\\
 \lambda^3 & \lambda^2 & 1\\
  \lambda^3 & \lambda^2 & 1 
\end{array}
\right) \;\;\; \mbox{and} \;\;\;
m_\nu \sim \left( \begin{array}{ccc}
1 & 1 & 1\\
1 & 1 & 1\\
1 & 1 & 1
\end{array}
\right) \; .
\end{equation} 

In figure~\ref{fig1} we display the success of the different charge assignments in describing correctly the lepton masses and mixing angles with respect to $\lambda$
as well as the probability distribution of the values of $\delta$ for a fixed value of $\lambda$, assuming the seesaw mechanism responsible for neutrino masses.

The realization of a model with an FN symmetry is simple. In particular, the breaking of $U(1)_{\mathrm{FN}}$ is easily engineered. Furthermore, an FN symmetry is also often used for the description of the quark sector, quark masses as well as mixing angles. Hence, such a symmetry can be suitable for both leptons and quarks. In addition, it has been shown that it can also be compatible with the
particle assignment in a grand unified theory. Results, similar to those obtained with an FN symmetry, can also be achieved in extra-dimensional models in which particles are localized
differently in the additional dimension(s).  For models with an FN symmetry see~\cite{U1FNother}.

\mathversion{bold}
\section{Theories with non-abelian $G_f$}
\mathversion{normal}

If a discrete non-abelian $G_f$ is broken to (non-trivial) residual symmteries $G_e$ in the charged lepton and to $G_\nu=Z_2 \times Z_2$
in the neutrino sector, lepton mixing mixing angles and the Dirac phase (up to $\pi$) can be fixed~\cite{Gfresidual}. $G_e$ is chosen in such a way that
the three lepton generations can be distinguished, while $G_\nu$ is always fixed to the maximal residual symmetry for three Majorana neutrinos that does
not lead to any constraints on their masses. The requirement that the charged lepton mass matrix $m_l$
should be invariant under $G_e$ determines the contribution $U_e$ of charged leptons to the Pontecorvo-Maki-Nakagawa-Sakata (PMNS) mixing matrix $U_{PMNS}$,
while the request that $m_\nu$ is invariant under $G_\nu$ fixes $U_\nu$. So, also the form of the PMNS mixing matrix $U_{PMNS}=U_e^\dagger U_\nu$ is given
by $G_e$ and $G_\nu$, up to possible permutations of rows and columns of $U_{PMNS}$, since lepton masses are not predicted from $G_f$, $G_e$ and $G_\nu$. Consequently,
the lepton mixing angles and the Dirac phase are determined up to these permutations of rows and columns of $U_{PMNS}$. Furthermore, the 
  columns of $U_e$ and $U_\nu$ can be re-phased, so that Majorana phases are in general not fixed. 
  
  One of the very first implementations of this approach 
 that leads to non-zero $\theta_{13}$ and non-maximal $\theta_{23}$ has been discussed in~\cite{dATFH11}. The choice of $G_f$ is $G_f=\Delta (96)$. The residual
 symmetries are $G_e=Z_3$ and $G_\nu=Z_2\times Z_2$ and lead to a PMNS mixing matrix whose elements have the absolute values
\begin{equation}
||U_{PMNS}|| =\frac{1}{\sqrt{3}}
\left(
\begin{array}{ccc}
\frac 12 \, \left( \sqrt{3} +1 \right) & 1 & \frac 12 \, \left( \sqrt{3} -1 \right)\\
 1 & 1 & 1\\
  \frac 12 \, \left( \sqrt{3} -1 \right) & 1 &  \frac 12 \, \left( \sqrt{3} +1 \right)
\end{array}
\right) \; .
\end{equation} 
The results for the lepton mixing angles are
\begin{equation}
\sin^{2} \theta _{12} =\sin ^{2} \theta _{23}  =\frac{8-2 \, \sqrt{3}}{13} \approx 0.349 \;\; \mbox{and} \;\; \sin ^{2} \theta _{13}=\frac{2-\sqrt{3}}{6}\approx 0.045 \; .
\end{equation}
The Dirac phase is predicted to be trivial, $\sin\delta=0$. A grand unified theory with $G_f=\Delta (96)$ has been constructed in~\cite{KLSD96model}. In several studies~\cite{leptonmixGf},
in particular in~\cite{GrimusFonseca}, series of $G_f$, possible choices of $G_e$ and $G_\nu$ and the resulting mixing patterns have been analyzed.
 It has been observed that  $\sin\delta=0$ follows, if the lepton mixing angles are in accordance with the experimental data. 

This approach can be combined with an FN symmetry so that a simultaneous
understanding of lepton mixing parameters as well as charged lepton masses becomes possible. 
For $G_f$ being discrete, the symmetry breaking scale can be as low as the electroweak scale or even larger than the scale of grand unification which offers great freedom in building models. 
 Explicit model realizations with a non-abelian discrete $G_f$ are discussed in e.g.~\cite{modelsGf}.

\mathversion{bold}
\section{Theories with non-abelian $G_f$ and CP}
\mathversion{normal}

In a scenario with a discrete non-abelian $G_f$ and a CP symmetry, in which both symmetries are broken to (non-trivial) residual groups, it becomes
possible to not only determine the lepton mixing angles and $\delta$, but also the Majorana phases $\alpha$ and $\beta$. The CP symmetry that is
imposed in the fundamental theory acts in general non-trivially on flavor space~\cite{GrimusRebelo95}, i.e. for a set of scalars $\phi_i$ that transform in the same way under 
the gauge symmetries (and form a multiplet of $G_f$) a CP transformation $X$ acts as
\begin{equation}
\phi_i \rightarrow X_{ij} \, \phi_j^\star \;\; \mbox{with} \;\; X X^\dagger = X X^\star = 1 \, .
\end{equation}
 In order to consistently combine $G_f$ and CP certain conditions have to be fulfilled~\cite{GfCPidea,GfCPothers}.
In the following examples all such conditions are fulfilled.
The approach for fixing lepton mixing angles and predicting leptonic CP phases, presented in~\cite{GfCPidea},
assumes $G_f$ and CP and as residual symmetries $G_e$ and $G_\nu$. While $G_e$ has to fulfill the same constraints as in the approach without CP, $G_\nu$ is chosen as the direct product of a $Z_2$ symmetry,
contained in $G_f$, and the CP symmetry. 
Thanks to the latter choice it becomes possible to also predict the Majorana phases. Furthermore, one real free parameter, which affects in general all lepton mixing parameters, is introduced in the PMNS mixing matrix, since $G_\nu$ is no longer the maximal residual symmetry $Z_2\times Z_2$. A consequence of this free real parameter is the possibility to obtain results for lepton mixing angles in agreement with experimental
data and, at the same time, to achieve non-trivial values of the Dirac phase $\delta$. 
The actual form of the PMNS mixing matrix in this approach is obtained from the contribution $U_e$ to lepton mixing from charged leptons, determined by $G_e$,
and the contribution $U_\nu$ from neutrinos, which is subject to $G_\nu=Z_2 \times CP$. It can be shown that $U_\nu$ can be written as $U_\nu=\Omega_\nu \, R (\theta) \, K_\nu$
and thus $U_{PMNS}$ reads
\begin{equation}
U_{PMNS}=U^\dagger_e \Omega_\nu R ( \theta ) K_\nu
\end{equation}
with $\Omega_\nu$ being determined by the CP transformation $X$ and the residual $Z_2$ flavor symmetry, $R (\theta)$ being a rotation in one plane through the free parameter $\theta$, $0 \leq \theta < \pi$, 
and $K_\nu$ a diagonal matrix with entries $\pm 1$ and $\pm i$. The latter is related to the request to achieve positive neutrino masses. Like the approach given in the
preceding section, also here lepton masses are unconstrained. Hence, all statements made hold up to possible permutations of rows and columns of the PMNS mixing matrix. 

One example that shows the 
predictive power of this approach has been discussed in~\cite{GfCPidea}. For $G_f=S_4$, $G_e=Z_3$ and $G_\nu=Z_2\times CP$ the PMNS mixing matrix is of the form
\begin{equation}
U_{PMNS} =\frac{1}{\sqrt{6}} \, \left( \begin{array}{ccc}
 2 \cos \theta & \sqrt{2} & 2 \sin \theta \\
 -\cos \theta + i \sqrt{3} \sin \theta & \sqrt{2} & -\sin \theta - i \sqrt{3} \cos \theta \\
 -\cos \theta - i \sqrt{3} \sin \theta & \sqrt{2} & -\sin \theta + i \sqrt{3} \cos \theta \\
\end{array}
\right) \; K_\nu
\end{equation}
which leads to lepton mixing angles
\begin{equation}
\sin ^{2} \theta _{13} = \frac{2}{3} \sin^2 \theta  \; , \;\;\; \sin^{2} \theta _{12}= \frac{1}{2 + \cos 2 \theta}  \; , \;\; \sin ^{2} \theta _{23}= \frac{1}{2}
\end{equation}
and CP phases
\begin{equation}
 |\sin \delta|=1 \;\; , \;\;\; \sin \alpha =0 \;\; \mbox{and} \;\;\; \sin \beta=0 \, .
\end{equation}
The Dirac phase is thus maximal, whereas both Majorana phases are trivial. Furthermore, the atmospheric mixing angle is fixed to be maximal.
The reactor and the solar mixing angle depend on the free parameter $\theta$ and for $\theta\approx 0.18$ or $\theta\approx 2.96$ both, $\theta_{13}$ and $\theta_{12}$,
are in agreement with experimental data. In~\cite{S4CPmodel} a supersymmetric model for the lepton sector with the gauge group of the Standard Model
has been constructed. In this model LH leptons are unified in a(n irreducible, faithful) triplet, whereas RH charged leptons are singlets of $S_4$. Both symmetries,
$S_4$ and CP, are broken spontaneously at a high energy scale. The above-estimated size of $\theta$, needed for achieving values of $\theta_{13}$ and $\theta_{12}$ 
consistent with experimental data, can be naturally explained in this model. Furthermore, neutrinos are predicted to follow normal mass ordering (NO) and the values of
the neutrino masses $m_i$ are 
\begin{equation}
m_1\approx 0.016 \, \mbox{eV} \; , \;\; m_2\approx 0.018 \, \mbox{eV} \; , \;\; m_3\approx 0.052 \, \mbox{eV} \, .
\end{equation}
In addition, the Majorana phases are fixed to the values $\alpha=\pi$ and $\beta=\pi$ so that $m_{ee}$, the quantity measurable in neutrinoless double beta decay, is 
$m_{ee}\approx 0.003 \, \mbox{eV}$. The charged lepton mass hierarchy is also naturally described, since charged lepton masses arise from operators of different dimension.

\begin{table}
\begin{center}
\begin{tabular}{c|ccc|cc}
\hline
$s$&$\sin^2 \theta_{13}$&$\sin^2 \theta_{12}$&$\sin^2 \theta_{23}$&$\sin\delta$&$\sin\alpha=\sin\beta$\\
\hline
$s=1$&$0.0220$&$0.318$&$0.579$&$0.936$&$-1/\sqrt{2}$\\
&$0.0220$&$0.318$&$0.421$&$-0.936$&$-1/\sqrt{2}$\\
\hline
$s=2$&$0.0216$&$0.319$&$0.645$&$-0.739$&$1$\\
\hline
$s=4$&$0.0220$&$0.318$&$0.5$&$\mp 1$&$0$\\
\hline
\end{tabular}
\end{center}
\caption{Results for lepton mixing parameters from $G_f=\Delta (6 \, n^2)$ with $n=8$, $m=4$ and different CP transformations $X (s)$. The matrix $K_\nu$ is chosen as trivial. 
The absolute value of $\sin\delta$ is large and the two Majorana phases $\alpha$ and $\beta$ take different values for different $s$.\label{tab1}}
\end{table}

In~\cite{DeltaCP} (see also~\cite{DeltaCPother}) the series $\Delta (3 \, n^2)$ and $\Delta (6 \, n^2)$ combined with CP have been analyzed in detail. The residual 
symmetries $G_e$ and $G_\nu$ are fixed to $G_e=Z_3$ and $G_\nu=Z_2\times CP$. It has been shown that for these types of residual symmetries only four cases of mixing patterns
can arise that lead to lepton mixing angles potentially compatible with experimental data. One particularly interesting case, called Case 3 b.1) in~\cite{DeltaCP}, has the following
features: the first column of the PMNS mixing matrix is fixed via the choice of the residual flavor symmetry $Z_2 (m)$ ($m$ integer); the solar mixing angle constrains $m$ to fulfill $m \approx n/2$; the free parameter $\theta$ is fixed by the reactor mixing angle and for $m=n/2$ a lower limit on the CP violation via the Dirac phase is found 
\begin{equation}
|\sin\delta| \gtrsim 0.71
\end{equation}
 and both Majorana phases $\alpha$ and $\beta$ depend on the CP transformation $X (s)$ only
 \begin{equation}
 |\sin\alpha|=|\sin\beta|=|\sin 6 \, \phi_s| \;\; \mbox{with} \;\; \phi_s=\frac{\pi s}{n} \;\; \mbox{and} \;\; s=0, ..., n-1 \, .
\end{equation} 
In table \ref{tab1} results for the lepton mixing parameters are shown for $\Delta (6 \, n^2)$ with $n=8$ and $m=4$ and different values $s$.

\begin{figure}
\parbox{6in}{\includegraphics[scale=0.33]{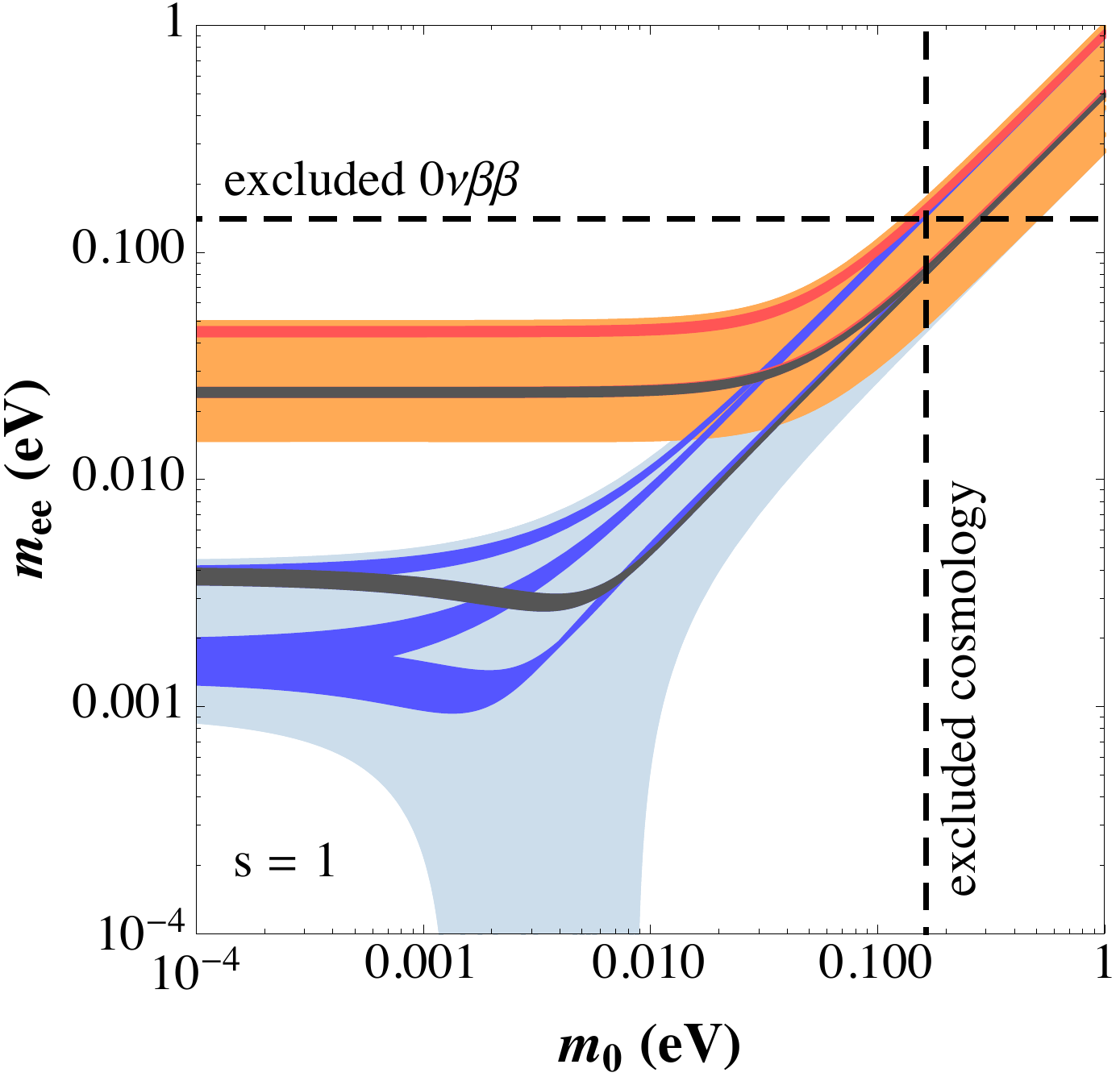}\hspace{0.1in}
\includegraphics[scale=0.33]{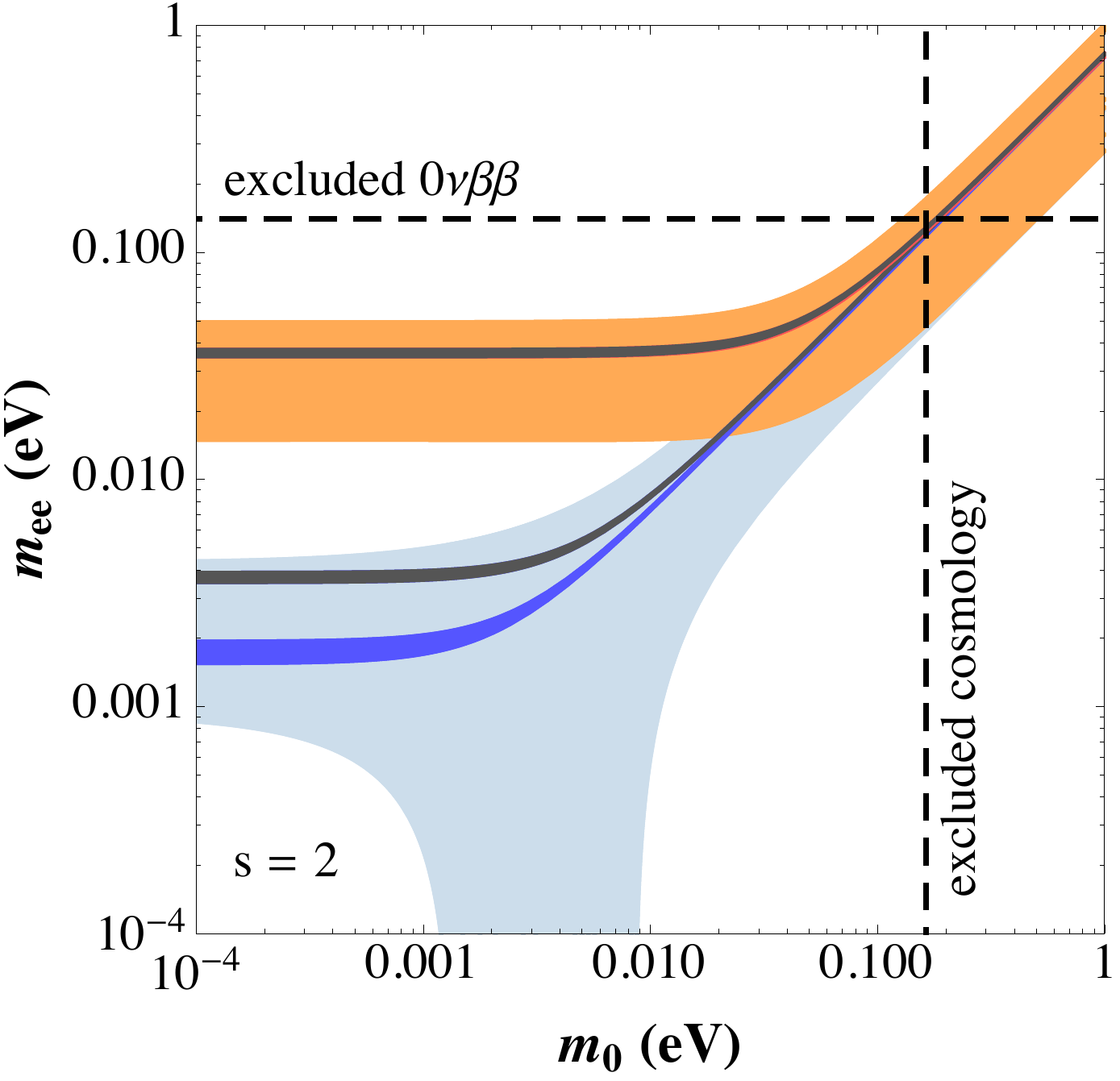}\hspace{0.1in}
\includegraphics[scale=0.33]{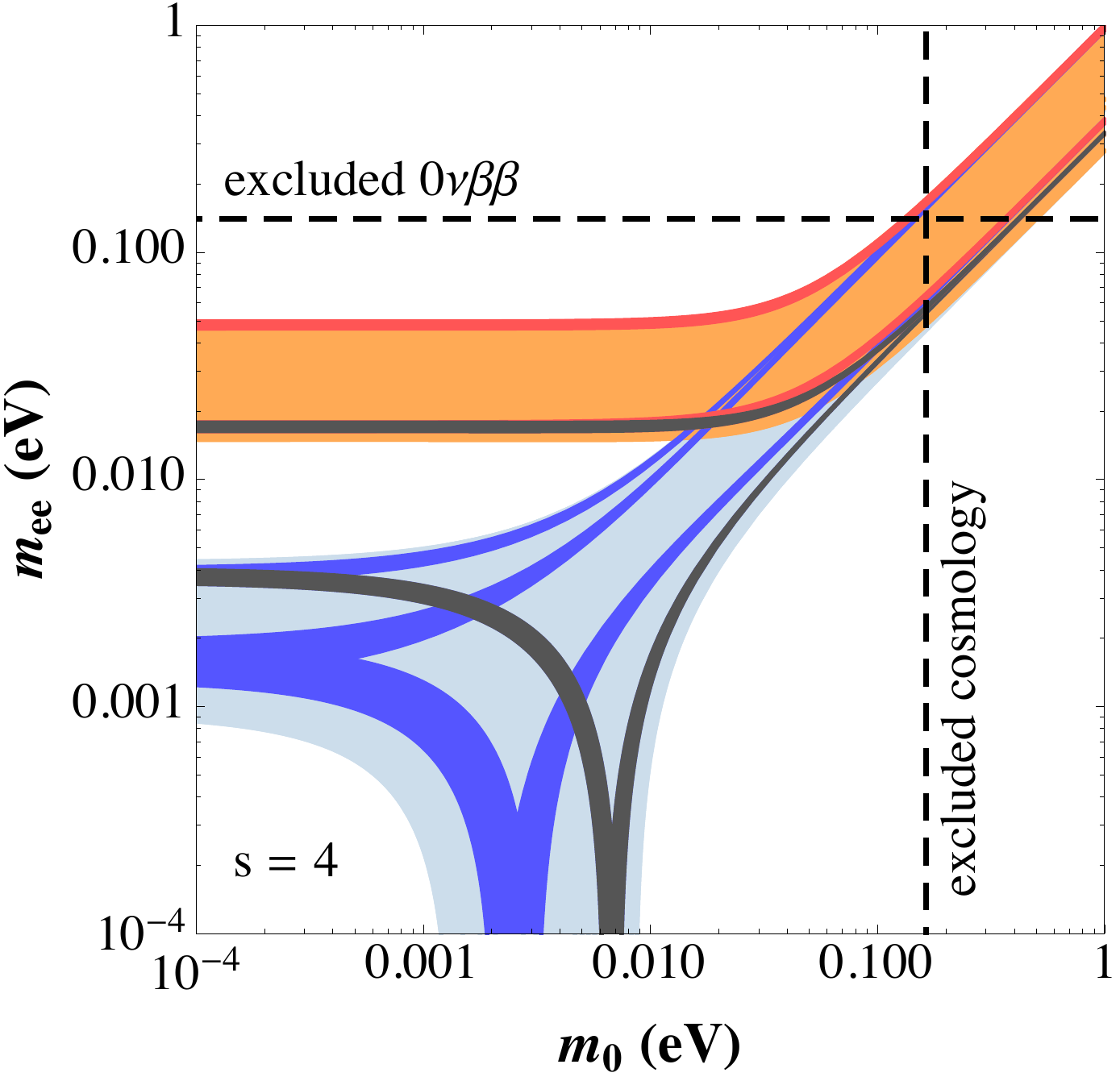}}
\caption{Results for $m_{ee}$ with respect to the lightest neutrino mass $m_0$ for the choices of $G_f$, CP, $G_e$ and $G_\nu$ used in table~\ref{tab1}.
Blue areas indicate $m_{ee}$ for NO, while orange areas refer to $m_{ee}$ for inverted mass ordering.
In dark colors the impact of the restrictions on the lepton mixing parameters on $m_{ee}$ is displayed, assuming for neutrino masses only the experimental constraints. 
For comparison in light colors the ranges of $m_{ee}$ are shown, as obtained from the experimentally preferred $3 \, \sigma$ intervals of the lepton mixing parameters and neutrino masses. 
 The darkest color highlights $K_\nu$ trivial. \label{fig2}}
\end{figure}

The fact that both, lepton mixing angles and Majorana phases, are strongly constrained leads also to strong restrictions on $m_{ee}$, even if the neutrino mass spectrum is not constrained
beyond the request to reproduce experimental bounds on the sum of the neutrino masses and to match the measured mass squared differences $\Delta m_{\mathrm{sol}}^2$ and $\Delta m_{\mathrm{atm}}^2$. This is exemplified in figure~\ref{fig2} for the choices of $G_f$, CP, $G_e$ and $G_\nu$ used in table~\ref{tab1}.

\begin{figure}
\parbox{6.5in}{\includegraphics[scale=0.35]{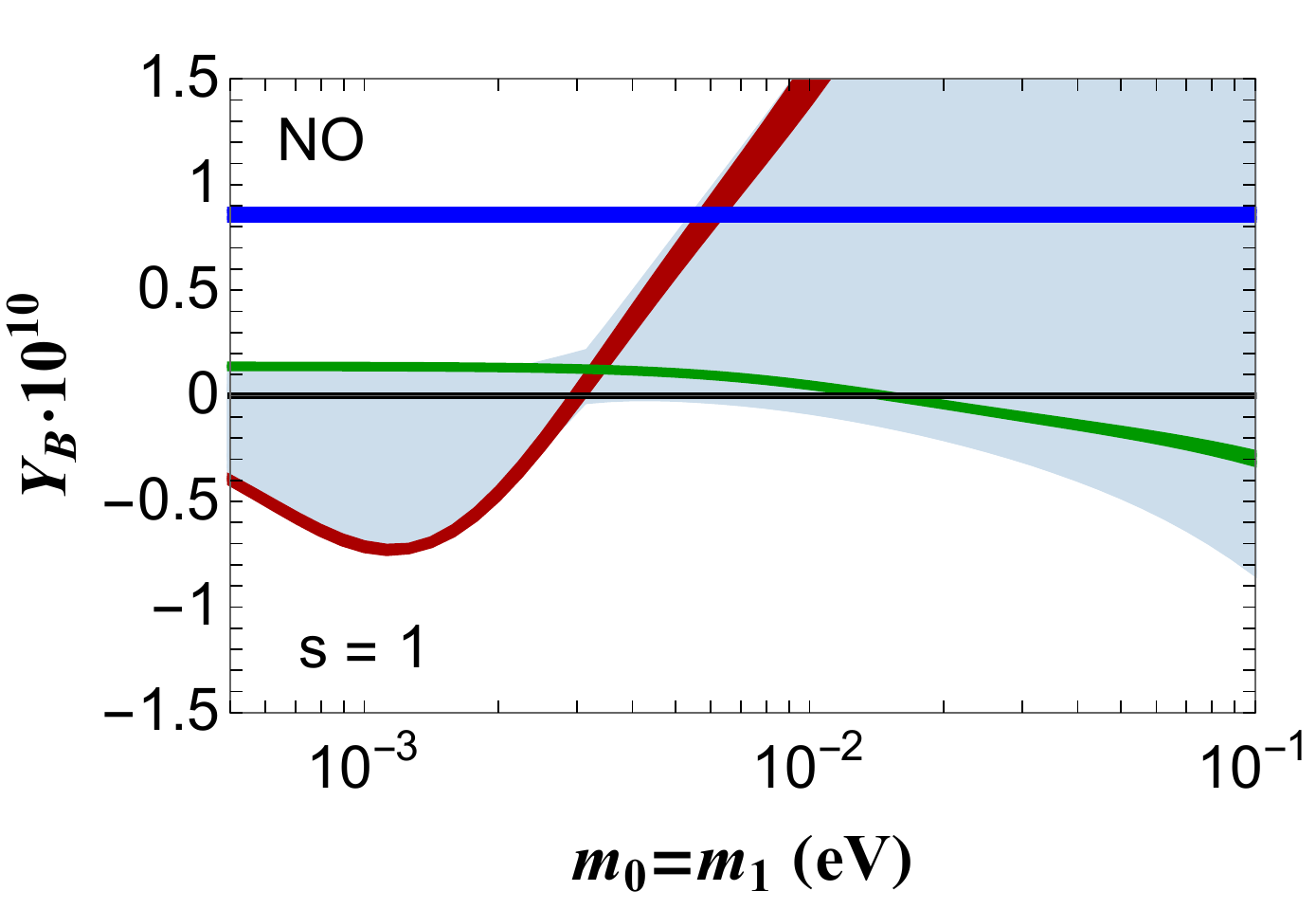}\hspace{0.1in}
\includegraphics[scale=0.35]{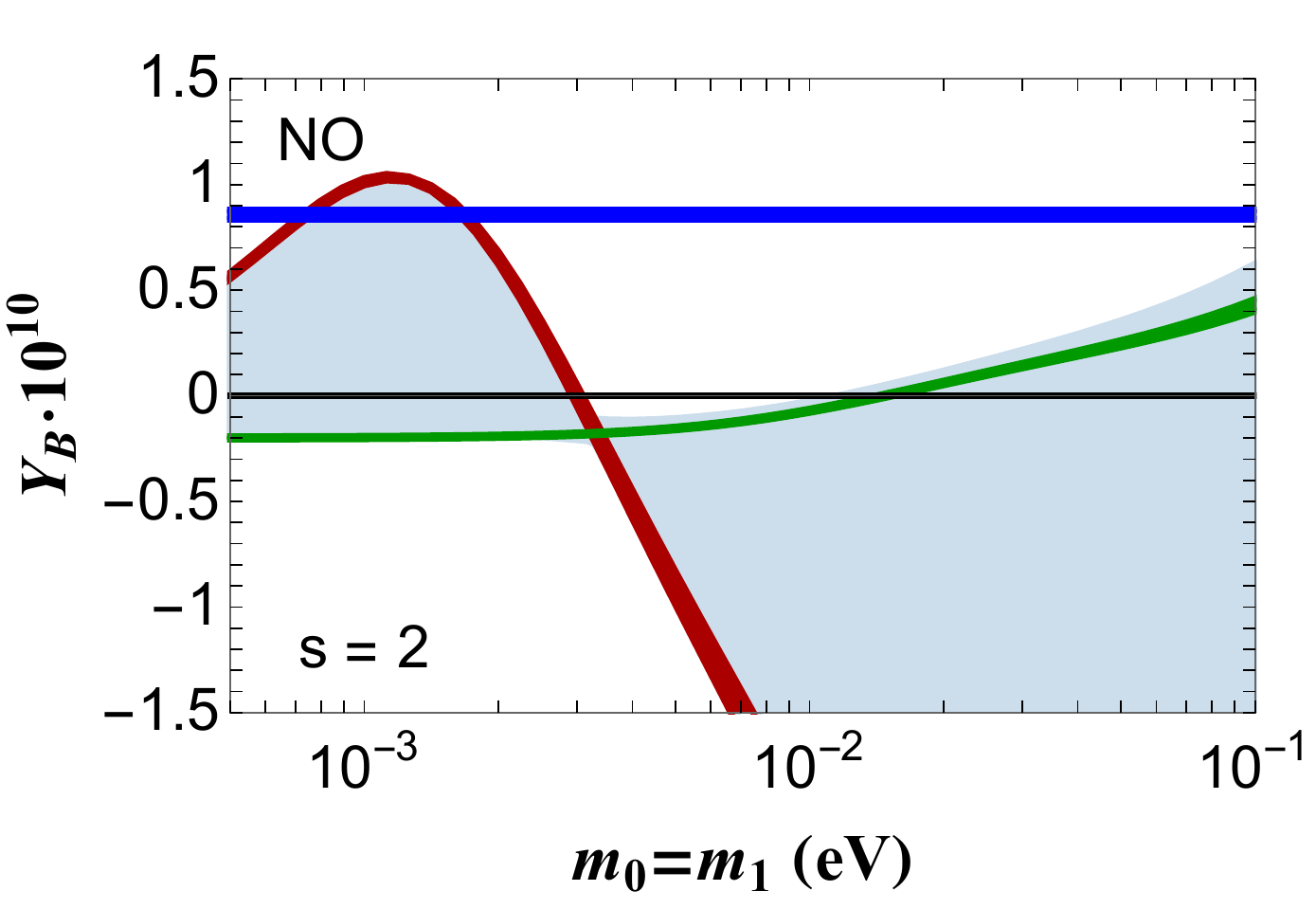}\hspace{0.1in}
\includegraphics[scale=0.35]{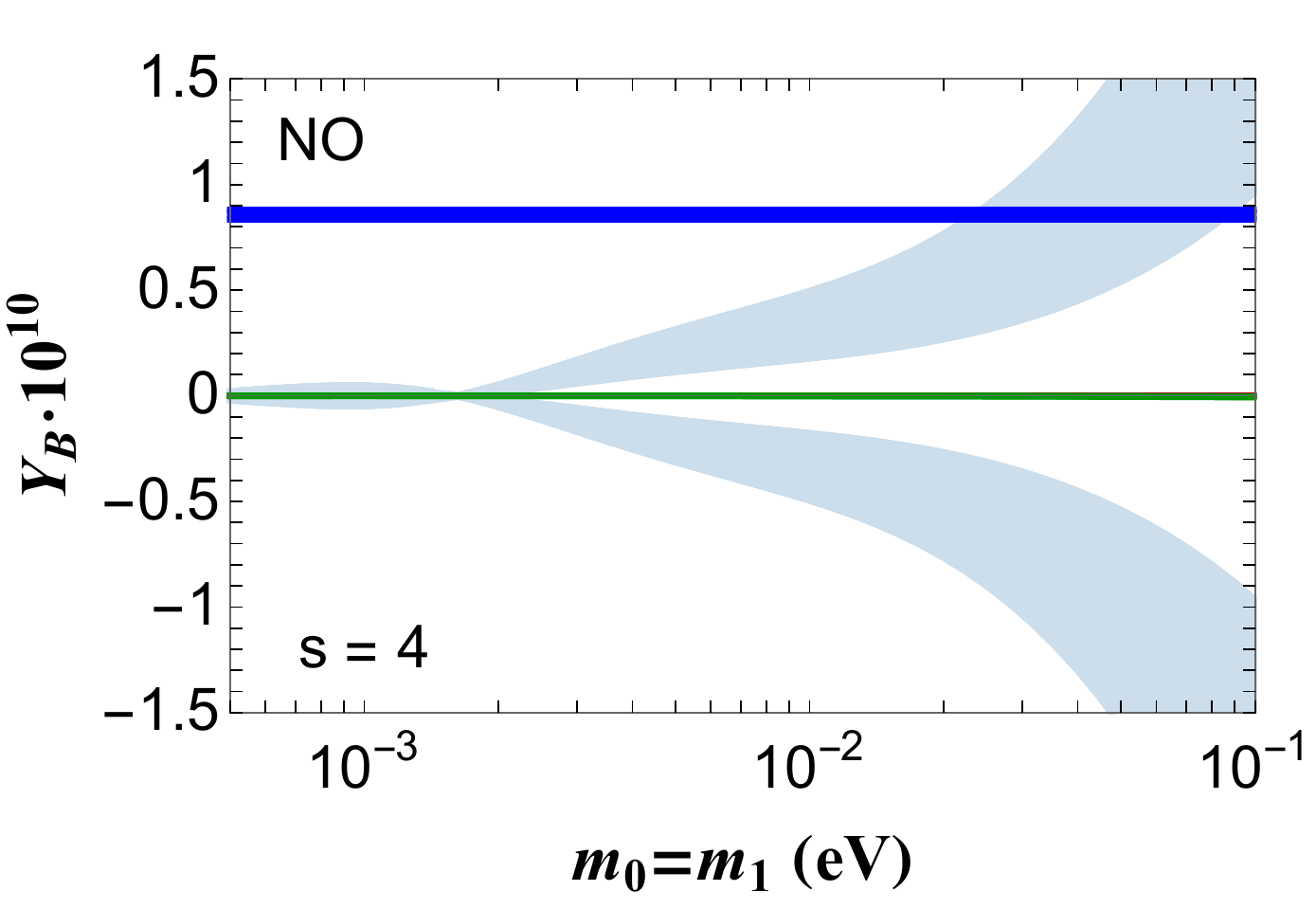}}
\caption{Results for the baryon asymmetry $Y_B$ of the Universe with respect to the lightest neutrino mass $m_0$ for the choice of $G_f$, CP, $G_e$ and $G_\nu$ 
as in table~\ref{tab1} and figure~\ref{fig2}. Light neutrino masses have NO. Light-blue, red and green areas refer to different choices of the parameters in the correction to $Y_D$.
The dark-blue area indicates the experimentally measured value of $Y_B$. For the choices $s=1$ and $s=2$ (predominantly) positive or negative $Y_B$ is achieved
for certain ranges of $m_0$. \label{fig3}}
\end{figure}

Given the predictive power of the approach with $G_f$ and CP regarding leptonic CP phases it has been applied in~\cite{GfCPlepto} to a scenario with three RH neutrinos $N_i$. 
$N_i$ transform in the same triplet as LH leptons $L_\alpha$. They give masses to light neutrinos via the type-I seesaw mechanism. 
For $10^{12} \, \mathrm{GeV} \lesssim M_i \lesssim 10^{14} \, \mathrm{GeV}$ the baryon asymmetry $Y_B$ of the Universe can be generated via unflavored leptogenesis ~\cite{FukugitaYanagida}, $Y_B \sim 10^{-3} \, \epsilon \, \eta$. A value of $Y_B$ in accordance with experimental data~\cite{Planck15}, $Y_B = (8.65 \pm 0.09) \times 10^{-11}$, can be achieved for CP asymmetries $10^{-4} \gtrsim \epsilon \gtrsim 10^{-7}$ for efficiency factors $10^{-3} \lesssim \eta \lesssim 1$.
In order to implement the breaking scheme of $G_f$ and CP, as described before, the charged lepton sector is taken to be invariant under $G_e$, while
the mass matrix $M_R$ of RH neutrinos preserves $G_\nu$ and the Dirac Yukawa coupling $Y_D$ is invariant under $G_f$ and CP. As a consequence, light neutrino
masses $m_i$ are inversely proportional to RH neutrino masses $M_i$ and the contribution $U_\nu$ from neutrinos to the PMNS mixing matrix is $U_\nu=U_R=\Omega_\nu R ( \theta ) K_\nu$.
Since charged leptons do not contribute to lepton mixing in the chosen basis, $U_{PMNS}=U_\nu$. Computing the CP asymmetries $\epsilon_i$, arising from the decay of $N_i$,
they are found to vanish. This has already been observed in scenarios with $G_f$ only~\cite{Gfonlylepto}. Thus, non-zero $\epsilon_i$ can be achieved, if corrections are included. A particularly
interesting case is that corrections to $Y_D$ are considered that are proportional to a (small) symmetry breaking parameter $\kappa$ and are invariant under $G_e$, the residual symmetry in the 
charged lepton sector. Taking these corrections into account, 
\begin{equation}
\epsilon_i \propto \kappa^2 \, . 
\end{equation}
Hence $\kappa \sim 10^{-(2\div 3)}$ explains correctly
the size of the CP asymmetries. Most importantly, the sign of $\epsilon_i$ (and consequently also $Y_B$) can be fixed, because all CP phases are determined in this approach.
In figure~\ref{fig3} the results for $Y_B$ as function of the lightest neutrino mass $m_0$ are shown. The light-blue, red and green areas arise from the variation of order one parameters appearing in the correction to $Y_D$. The choice of $G_f$, CP, $G_e$ and $G_\nu$ is the same as in table~\ref{tab1} and figure~\ref{fig2}.
As can be clearly seen, for certain choices of CP, $s=1$ and $s=2$, and certain ranges of $m_0$, $Y_B$ is (predominantly) positive or negative, whereas for the choice $s=4$ no such preference is visible. The explanation for this observation is that for $s=1$ the Majorana phase $\alpha$ fulfills $\sin\alpha<0$, whereas for $s=2$ we find $\sin\alpha>0$. For $s=4$ the CP phases $\alpha$
 and $\beta$ are trivial and only $\sin\delta$ is non-vanishing.  Studies of flavored leptogenesis in scenarios with $G_f$ and CP can be found in~\cite{GfCPflavorlepto}.

\section{Conclusions}

I have discussed for different flavor symmetries $G_f$ (abelian and non-abelian, continuous and discrete, combined with CP or not) 
their predictive power regarding lepton masses and lepton mixing parameters, in particular leptonic CP phases. While an FN symmetry is suitable
for (charged lepton) mass hierarchies and for explaining the gross structure of the mixing pattern, non-abelian $G_f$, especially if chosen to be discrete and broken
non-trivially, can explain all three lepton mixing angles and the Dirac phase $\delta$. However, their predictive power regarding CP phases is limited, since only one CP phase can be determined.
 A combination of non-abelian discrete $G_f$ 
and CP is most powerful in constraining all lepton mixing parameters and can also restrict high energy CP phases that are relevant
for the baryon asymmetry $Y_B$ of the Universe  in leptogenesis scenarios. I have also briefly shown that in concrete models the predictive power can be further increased, e.g. 
 the neutrino mass ordering is predicted and the Majorana phases are entirely fixed.

\Acknowledgements

I would like to thank the organizers of NuPhys2016 for the opportunity to give a talk as well as my collaborators with whom I worked on the 
different presented projects.


\end{document}